\documentclass{ws-procs9x6}
\def\be{\begin{equation}}
\def\ee{\end{equation}}
\def\bg{\begin{eqnarray}}
\def\en{\end{eqnarray}}

\begin{document}

\title{Binding of hypernuclei, and phtoproduction of $\Lambda$-hypernuclei\\
in the latest quark-meson coupling model}

\author{K. Tsushima$^1$, P.~A.~M.~Guichon$^{2}$, R.~Shyam$^{1,3}$, A.~W.~Thomas$^{1,4}$}

\address{$^1$Thomas Jefferson Lab., 12000 Jefferson Ave., Newport News, VA 23606, USA}
\address{$^{2}$ SPhN-DAPNIA, CEA Saclay, F91191 Gif sur Yvette, France}
\address{$^{3}$ Saha Institute of Nuclear Physics, Kolkata 70064, India}
\address{$^{4}$ College of William and Mary, Williamsburg VA 23187, USA}

\begin{abstract}
We study the binding of hypernuclei based on the latest version of
quark-meson coupling model, and estimate the phtoproduction
cross sections for the $^{12}$C($\gamma,K^+$)$^{12}_\Lambda$B reaction using
the bound $\Lambda$ spinors obtained in the model.
\end{abstract}


\bodymatter

\section{Introduction}\label{intro}

The study of $\Lambda$ hypernuclei has provided us with important information
on the properties of $\Lambda$ in a nuclear medium and the effective $\Lambda$-N
interaction~\cite{Hashimoto:2006aw}. On the other hand, the situation for $\Sigma$ and $\Xi$
hypernuclei is quite different.
The special case of $^4$He aside, there is no experimental
evidence for any $\Sigma$ hypernuclei~\cite{Saha:2004ha},
despite extensive searches.
It seems likely that the $\Sigma$-nucleus interaction is somewhat repulsive and that
there are no bound $\Sigma$ hypernuclei beyond A=4. In the case of the $\Xi$,
the experimental situation is very challenging, but we eagerly await studies of
$\Xi$ hypernuclei with new facilities at J-PARC and GSI-FAIR.

To understand further the properties of hypernuclei, we apply the latest version of
the quark-meson coupling (QMC) model~\cite{RikovskaStone:2006ta}, and calculate
the single-particle energies~\cite{qmcGTT} and phtoproduction cross sections~\cite{qmcphotohyp}.
The major improvement in the latest version is the inclusion of the effect
of the medium on the color-hyperfine interaction. This has the effect of increasing the splitting
between the $\Lambda$ and $\Sigma$ masses as the density rises. This is the prime
reason why our results yield no middle and heavy mass $\Sigma$ hypernuclei~\cite{qmcGTT}.

The QMC model was created to provide insight into the structure of nuclear matter,
starting at the quark level~\cite{Guichon:1987jp,qmcPPNPreview}.
Nucleon internal structure was modeled using the MIT bag,
while the binding was described by the self-consistent coupling of the
confined quarks to the scalar-$\sigma$ and vector-$\omega$ meson fields generated by the confined
quarks in the other ``nucleons'' in the medium.
The self-consistent response of the bound quarks to the mean $\sigma$ field leads
to a novel saturation mechanism for nuclear matter, with the enhancement of
the lower components of the valence Dirac wave functions.
The model has been successfully used to study
various nuclear phenomena~\cite{qmcPPNPreview}.

\section{Results for hypernuclei}\label{yspectra}

To calculate the hyperon levels, we use a relativistic shell model.
Details on the calculations are described in Refs.~\cite{qmcGTT,qmchyp}.
Results for the hypernuclear single-particle energies
are shown in Tables~\ref{spe1} and~\ref{spe2}.

First, we emphasize that the present calculation yields no $\Sigma$ hypernuclei in
the nuclei considered. This is consistent with the empirical fact of no finding of the
middle and heavy mass $\Sigma$ hypernuclei.

Second, overall agreement with the experimental energy levels of $\Lambda$ hypernuclei
across the periodic table is quite good. In particular, the parameter free result of
-26.9 MeV for the $1s_{1/2}$ level of $^{208}$Pb is impressive.
The discrepancies which remain may well be resolved by small effective
hyperon-nucleon interactions which go beyond the simple, single-particle shell model.
Once again, we stress the very small spin-orbit force experienced by the $\Lambda$,
which is a natural property of the QMC model~\cite{qmchyp}.

Finally, this model supports the existence of a variety of bound $\Xi$-hypernuclei.
For the $\Xi^0$ the binding of the $1s_{1/2}$ level varies from 5 MeV in $^{17}_{\Xi^0}$O to
15 MeV in $^{209}_{\Xi^0}$Pb. The experimental search for such states at facilities such as
J-PARC and GSI-FAIR will be very important.

\begin{table}[htbp]
\tbl{Single-particle energies (in MeV)
for $^{17}_Y$O, $^{41}_Y$Ca and $^{49}_Y$Ca
hypernuclei.
The experimental data are taken from
Ref.~\protect\cite{Hashimoto:2006aw} (Table 11) for
$^{16}$O and from Ref.~\protect\cite{Pile:1991cf} for $^{40}$Ca. }
{\begin{tabular}{c|ccc|ccc|cc}
\hline
\hline
&$^{16}_\Lambda$O(Exp.) &$^{17}_\Lambda$O &$^{17}_{\Xi^0}$O
&$^{40}_\Lambda$Ca(Exp.)&$^{41}_\Lambda$Ca &$^{41}_{\Xi^0}$Ca
&$^{49}_\Lambda$Ca     &$^{49}_{\Xi^0}$Ca\\
\hline
\hline
$1s_{1/2}$&-12.42 &-16.2 &-5.3 &-18.7 &-20
.6 &-5.5 &-21.9 &-9.4 \\
&$\pm0.05$& & &$\pm 1.1$& & & & \\
&$\pm0.36$& & & & & & & \\
$1p_{3/2}$&           & -6.4 &---  &            &-13.9 &-1.6 &-15.4 &-5.3 \\
$1p_{1/2}$& -1.85 & -6.4 &---  & &-13.9 &-1.9 &-15.4 &-5.6\\
&$\pm 0.06$& & & & & & & \\
&$\pm 0.36$& & & & & & & \\
\end{tabular}}\label{spe1}
\end{table}
\begin{table}[htbp]
\tbl{Same as table~\ref{spe1} but
for $^{91}_Y$Zr and $^{208}_Y$Pb hypernuclei. The experimental data are
taken from Ref.~\protect\cite{Hashimoto:2006aw} (Table 13).}
{\begin{tabular}{c|ccc|ccc}
\hline
\hline
&$^{89}_\Lambda$Yb(Exp.)  &$^{91}_\Lambda$Zr  &$^{91}_{\Xi^0}$Zr
&$^{208}_\Lambda$Pb(Exp.) &$^{209}_\Lambda$Pb &$^{209}_{\Xi^0}$Pb \\
\hline
\hline
$1s_{1/2}$&-23.1 $\pm 0.5$       &-24.0 &-9.9 &-26.3 $\pm 0.8$&-26.9 &-15.0 \\
$1p_{3/2}$&            &-19.4 &-7.0 &            &-24.0 &-12.6 \\
$1p_{1/2}$&-16.5 $\pm 4.1$ ($1p$)&-19.4 &-7.2 &-21.9 $\pm 0.6$ ($1p$)&-24.0 &-12
.7 \\
$1d_{5/2}$&            &-13.4 &-3.1 &---         &-20.1 & -9.6 \\
$2s_{1/2}$&            & -9.1 &---  &---         &-17.1 & -8.2 \\
$1d_{3/2}$&-9.1 $\pm 1.3$  ($1d$)&-13.4 &-3.4 &-16.8 $\pm 0.7$ ($1d$)&-20.1 & -9
.8 \\
$1f_{7/2}$&            & -6.5 &---  &---         &-15.4 & -6.2 \\
$2p_{3/2}$&            & -1.7 &---  &---         &-11.4 & -4.2 \\
$1f_{5/2}$&-2.3 $\pm1.2$  ($1f$)& -6.4 &---  &-11.7 $\pm 0.6$ ($1f$)&-15.4 & -6.
5 \\
$2p_{1/2}$&            & -1.6 &---  &---         &-11.4 & -4.3 \\
$1g_{9/2}$&            &---   &---  &---         &-10.1 & -2.3 \\
$1g_{7/2}$&            &---   &---  & -6.6 $\pm 0.6$ ($1g$)&-10.1 & -2.7
\end{tabular}}\label{spe2}
\end{table}

\begin{figure}[htbp]
\begin{center}
\psfig{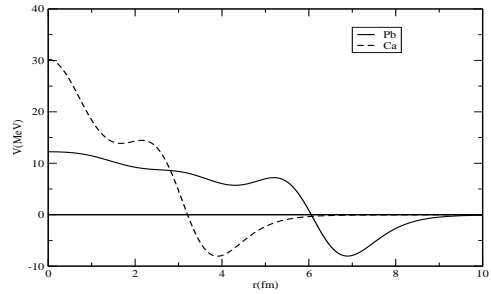}
\caption{$\Sigma^0$ potential in $^{40}$Ca and $^{208}$Pb nuclei.}
\label{fig:pot}
\end{center}
\end{figure}

Concerning the absence of the $\Sigma$ hypernuclei in the present calculation,
it is especially interesting to examine the effective
non-relativistic potential felt by the $\Sigma^0$ in a finite nucleus.
This is shown in Fig.~\ref{fig:pot} for $^{40}$Ca and $^{208}$Pb nuclei.
In the central region the vector interaction dominates over the scalar one leading
to a repulsive effective potential which reaches respectively 30 MeV and 12 MeV
at the center. It is only at the surface that the scalar attraction   becomes dominant.

\section{Photoproduction of $\Lambda$-hypernuclei}\label{photohyp}

In existing several theoretical studies of photoproduction of hypernuclei,
it has been usually employed nonrelativistic models to obtain the relevant initial and final state
wave functions (except for Ref.~\cite{ben89}).
In this study, we follow Ref.~\cite{shy08} and use a fully covariant model
to calculate the cross sections for the $^{12}$C ($\gamma,K^+) ^{12}_\Lambda$B reaction.
We explore the feasibility of studying the photoproduction
of hypernuclei within the relativistic model of Ref.~\cite{shy08}, but
employing the bound $\Lambda$ spinors obtained by the latest
QMC model~\cite{qmcGTT}. This provides an opportunity to investigate the role
of the quark degrees of freedom in the hypernuclear production, which may be
a novel feature.

The relevant processes included in the present study are shown in
Figs.~1(a) and 1(b), the elementary $\gamma p \to K^+ \Lambda$ and
the hypernuclear production reactions, $A (\gamma,K^+) _{\Lambda}B$,
respectively. In principle, although the $u$- and $t$-channels
should also be included, they contribute to the
non-resonant background terms which are insignificant to both elementary
as well as in-medium photon induced reactions for energies
below 1.5 GeV. (See, e.g., Refs.~\cite{lee01,shy08s}.)
We use plane waves to describe the relative motion of the outgoing particle which
is justified by the relatively weaker kaon-nucleus interaction in the final channel.

\begin{figure}[htbp]
\begin{center}
\psfig{file=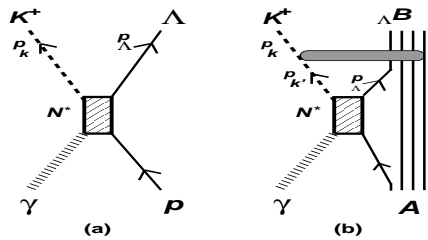,width=2.2in,height=1.2in}
\caption{Processes included in the calculation.}
\label{Sigma0pot}
\end{center}
\end{figure}

All ingredients necessary in the calculation, such as effective Lagrangians and
resonance propagators, are described in Refs.~\cite{shy08,shy04}.
The coupling constants have been determined by comparing our calculations [graph 1(a)] with the total
and differential cross section data for the elementary $\gamma p \to \Lambda K^+$ reaction
in the relevant photon energy region~\cite{qmcphotohyp}.

\begin{figure}[!t]
\begin{center}
\psfig{file=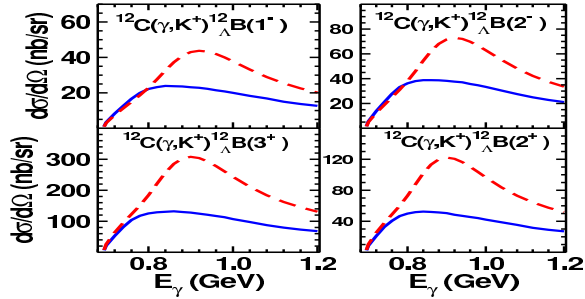,width=3in,height=1.5in}
\caption{ Differential cross sections (for the outgoing kaon
angle of 10$^\circ$) for the $^{12}$C$(p,K^+)$$^{12}_\Lambda$B reaction~\cite{qmcphotohyp}
leading to hypernuclear states. The solid and dashed lines show
the results of QMC and phenomenological models, respectively.}
\label{dsigma}
\end{center}
\end{figure}

The threshold for the kaon photoproduction on $^{12}$C is about 695 MeV.
The momentum transfer involved in the reaction at 10$^\circ$ kaon
angle, in which we focus, varies between approximately 2 fm$^{-1}$ to 1.4 fm$^{-1}$ in the
photon energy range of 0.7 GeV to 1.2 GeV~\cite{shy08e}.
In Fig.~\ref{dsigma}, we compare the differential cross section calculated~\cite{qmcphotohyp} with the
bound $\Lambda$ spinors obtained in the latest QMC model~\cite{qmcGTT}, and the
phenomenological models~\cite{shy08} for the $^{12}$C ($\gamma,K^+) ^{12}_\Lambda$B
reaction. The hole state spinor is taken from the phenomenological model~\cite{shy08} in
both cases.

The cross sections are shown for photon energies in the
range of  0.7-1.2 GeV corresponding to the outgoing kaon
angle of 10$^\circ$. The hypernuclear states populated are $1^-$, $2^-$,
and $2^+$, $3^+$ corresponding to the  particle-hole configurations of
$(1p_{3/2}^{-p},1s_{1/2}^\Lambda)$ and $(1p_{3/2}^{-p},1p_{3/2}^\Lambda)$,
respectively. We see that in each case the QMC cross sections are smaller
than those obtained with phenomenological hyperon spinors.
In this figure we further note that the peaks of the
QMC cross sections are somewhat shifted toward lower photon energies as
compared to those of the phenomenological model.
Detailed explanations for these features are given in Ref.~\cite{qmcphotohyp}.

We further note that within each group the highest $J$ state is most
strongly excited, which is in line with the results presented in
Refs.~\cite{ben89,ros88,shy08}. Furthermore, unnatural parity states within
each group are preferentially excited by this reaction. The unnatural parity
states are excited through the spin flip process. Thus, this confirms
that kaon photo- and also electro-production reactions on nuclei are ideal
tools for investigating the structure of unnatural parity hypernuclear states.

\section{summary}

We have studied the properties of hypernuclei using the latest version of the
quark-meson coupling model, which includes the effect of the medium on the color-hyperfine interaction
between quarks. This latest version leads to some important results:
(1) The agreement between the parameter free calculations and the experimental ground state levels for $\Lambda$-hypernuclei from Calcium to Lead is impressive.
(2) A number of $\Xi$-hypernuclei are predicted to be bound,
although not as deeply as in the $\Lambda$ case.
(3) The additional repulsion arising from the enhancement of the color-hyperfine
interaction in the $\Sigma$-hyperon in-medium (together with the effect of the $\Sigma N -
\Lambda N$ channel coupling on the intermediate range scalar attraction~\cite{qmcGTT,qmchyp})
yields to predict "no bound" $\Sigma$-hypernuclei.

We have also studied photoproduction of hypernucleus by
the $^{12}$C ($\gamma,K^+) ^{12}_\Lambda$B reaction within a covariant model,
using the bound $\Lambda$ spinors obtained by the latest quark-meson coupling model.
This is the first time that quark degrees of freedom has been explicitly
invoked in the description of the hypernuclear production.

\section*{Acknowledgments}
K.T. would like to thank the organizers of SENDAI08, O. Hashimoto, H. Tamura,
S.~N. Nakamura and K. Maeda for a warm hospitality in entire period of the conference.
Notice: Authored by Jefferson Science Associates, LLC under U.S. DOE Contract No. DE-AC05-06OR23177. The U.S. Government retains a non-exclusive, paid-up, irrevocable, world-wide license to publish or reproduce this manuscript for U.S. Government purposes.


\end{document}